\documentclass{ws-procs9x6}

\newcommand{\rr}[4]{#1, {\it #2 \/}{\bf #3} #4}
 
\begin{document}

\title{BEYOND BFKL}

\author{R.~B. PESCHANSKI}

\address{CEA/DSM/SPhT,Unit\'e de recherche 
associ\'ee 
au CNRS, \\
CE-Saclay, F-91191 Gif-sur-Yvette Cedex, France; \\Email: 
pesch@spht.saclay.cea.fr}


\maketitle

\abstracts{The Balitsky-Fadin-Kuraev-Lipatov (BFKL) evolution equation is 
known 
to be  ``unstable'' with respect to  fluctuations in gluon virtuality, 
transverse 
momentum and energy requiring to go beyond the leading order BFKL.  Still, 
these instabilities point to   fruitful  improvements of our deep 
understanding of 
QCD. Recent applications to next-leading order  and to saturation problems are 
outlined.}

\section{``Instabilities'' of the BFKL Equation}
\begin{figure}[ht]
\centerline{\epsfxsize=3.9in\epsfbox{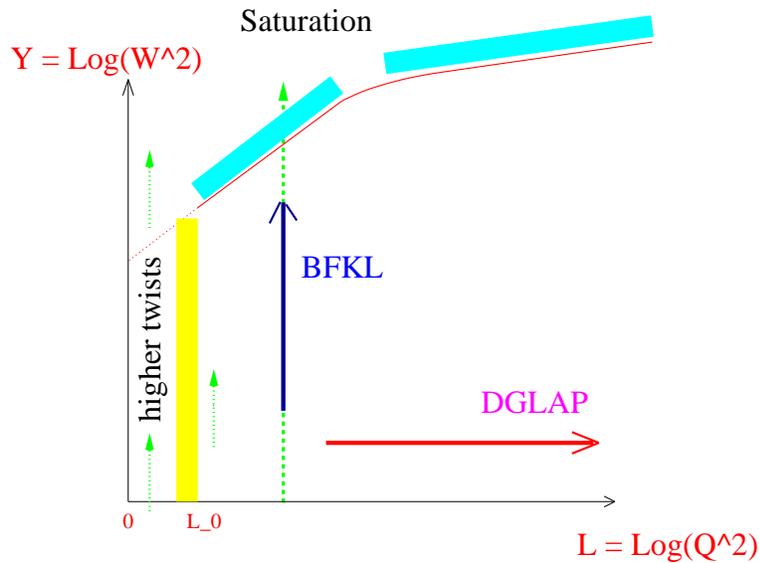}}   
\caption{{\it The QCD evolution landscape.} The BFKL evolution  and its 
limitations. {\bf i)} to the left: non-perturbative region; {\bf ii)} to the 
right: DGLAP evolution; {\bf iii)}  to the top: Saturation region.
 \label{map}}
\end{figure}
The Balitsky-Fadin-Kuraev-Lipatov (BFKL) evolution equation has already a 
venerable past\cite{bfkl}.
It appears as a key tool in many recent works on small-x physics (in the broad 
sense). It is interesting  to notice that  
its limitations themselves are the seeds of interesting fields of research.
Let 
us discuss  limitations which can be associated with the idea of 
``instabilities''.

 \begin{itemize}
\item 
{\bf i)} {\it Instability towards the non-perturbative regime}

It is well known that the perturbative ``gluon ladder'' contributing to the 
BFKL 
cross-section is characterized by a ``cigar-shape'' structure\cite{bartel}  of 
the transverse momenta. Hence, it is difficult to avoid an excursion 
inside the near-by non-perturbative region 

(Fig. (\ref{map}), to the 
left).

\item 
{\bf ii)} {\it Instability towards the renormalization group  regime}

Calculations of the next-leading BFKL kernel\cite{next} has proven that the
inclusion of next-leading logs gives a (too) strong correction to the leading 
log result. After a resummation motivated by the suppression of   spurious 
singularities\cite{salam,brod}, the results show that the resummed NLO-BFKL 
kernels are very similar, {\it e.g.} ``attracted'' towards the Dokshitzer 
Gribov 
Lipatov Altarelli Parisi (DGLAP)
 evolution\cite{dglap} , at least for the structure functions
 
  (Fig. (\ref{map}), to the 
right). 

\item 
{\bf iii)} {\it Instability towards the high density (saturation) regime}

The BFKL evolution implies a densification of gluons and 
sea quarks, while they keep in average the same size. It is thus natural to 
expect\cite {GLR} a modification of the evolution equation by non-linear 
contributions in the gluon density. Recently, the corresponding theoretical 
framework  has been settled\cite{venugopalan,balitsky}, and is based on an 
extension of the BFKL kernel acting on  non-linear terms. It  leads to a 
transition to the saturation regime
 (Fig. (\ref{map}), to the top). 

\end{itemize}

The main subjects of my talk will concern 
contributions\footnote{We shall leave 
thepoint 
({\bf i)} outside of the scope of the present conference, despite 
some recent  developments related to the AdS/CFT 
correspondence\cite{ma} and the ``BFKL treatment'' of the  4-Supersymmetrical 
gauge field theory\cite{strings}.}
to point ({\bf ii)}, with a 
discussion of the phenomenological relevance of  (resummed) NLO-BFKL kernels 
and  point ({\bf iii)}, with a discussion of traveling wave  solutions of 
non-linear QCD equations, as being deeply related to geometric scaling and the 
transition to saturation.

\section{ ``Instability towards DGLAP''}

The ``instability'' of the BFKL equation's solution w.r.t. the renormalization 
group evolution is well-known\cite{brux}. Indeed, the first correction to the 
leading-$\log 1/x$ approximation of the BFKL kernel\cite{next} 
is large enough to apparently endanger  the whole BFKL approach. It was soon 
realized that  a 
large 
part of the problem was due to the appearance of singularities which 
contradict 
the renormalization group  properties. Hence requiring  an harmonization 
between 
the next leading log BFKL calculations and the renormalization group 
requirements through  higher orders'  resummation leads\cite{salam,brod} to a 
possible way out of the problem.

Let us  focus\cite{us} on the  impact of these developments on the   proton 
structure functions
\begin{figure}[ht]
\begin{center} 
\epsfxsize=9cm  
\hspace{1.7cm}
\epsfbox{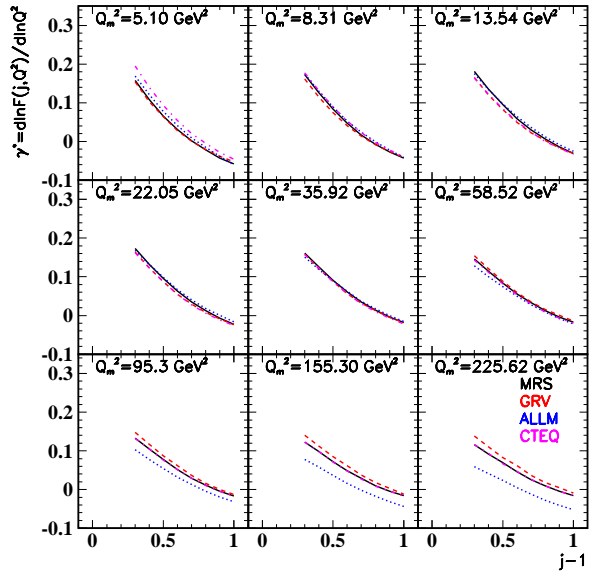} 
\vspace{-2cm} 
\caption{{\it  $\bar \gamma(\omega,Q^2):$ The ``effective'' $F_2$ anomalous 
dimension. }
$\bar \gamma(\omega,Q^2)$ has been evaluated from four known different 
parametrizations. They  are all compatible in the range
 $.3<\omega<1, 5<Q^2<100\ GeV^2,$ where we restrict our analysis. }
\end{center}
\label{fig2}
\end{figure}
and recall the parametrization of the proton structure functions in the 
(LO) BFKL 
approximation\cite{old}:
 \begin{equation}
F_i=
\int\frac{d \gamma}{2i\pi}\left(\frac{Q^2}{Q_0^2}\right)^{ \gamma} 
{x_{Bj}}^{-\frac {\alpha_s N_c}{\pi} \chi_{L0}( \gamma)}
\ h_i(\gamma)\ \eta( \gamma)\ ,
\label{BFKL}
\end{equation}
where $F_i$ denotes respectively  $F_T, \ F_L, \ G$ (resp. transverse, 
longitudinal and gluon) structure functions and $h_i$ are the known 
perturbative 
couplings to the photon ($h_G\!=\!1$ for the gluon structure function), 
usually 
called 
``impact factors''\cite{impact}.
\begin{figure}[ht]
\begin{center}
\epsfxsize=9cm   
\hspace{1.7cm}
\epsfbox{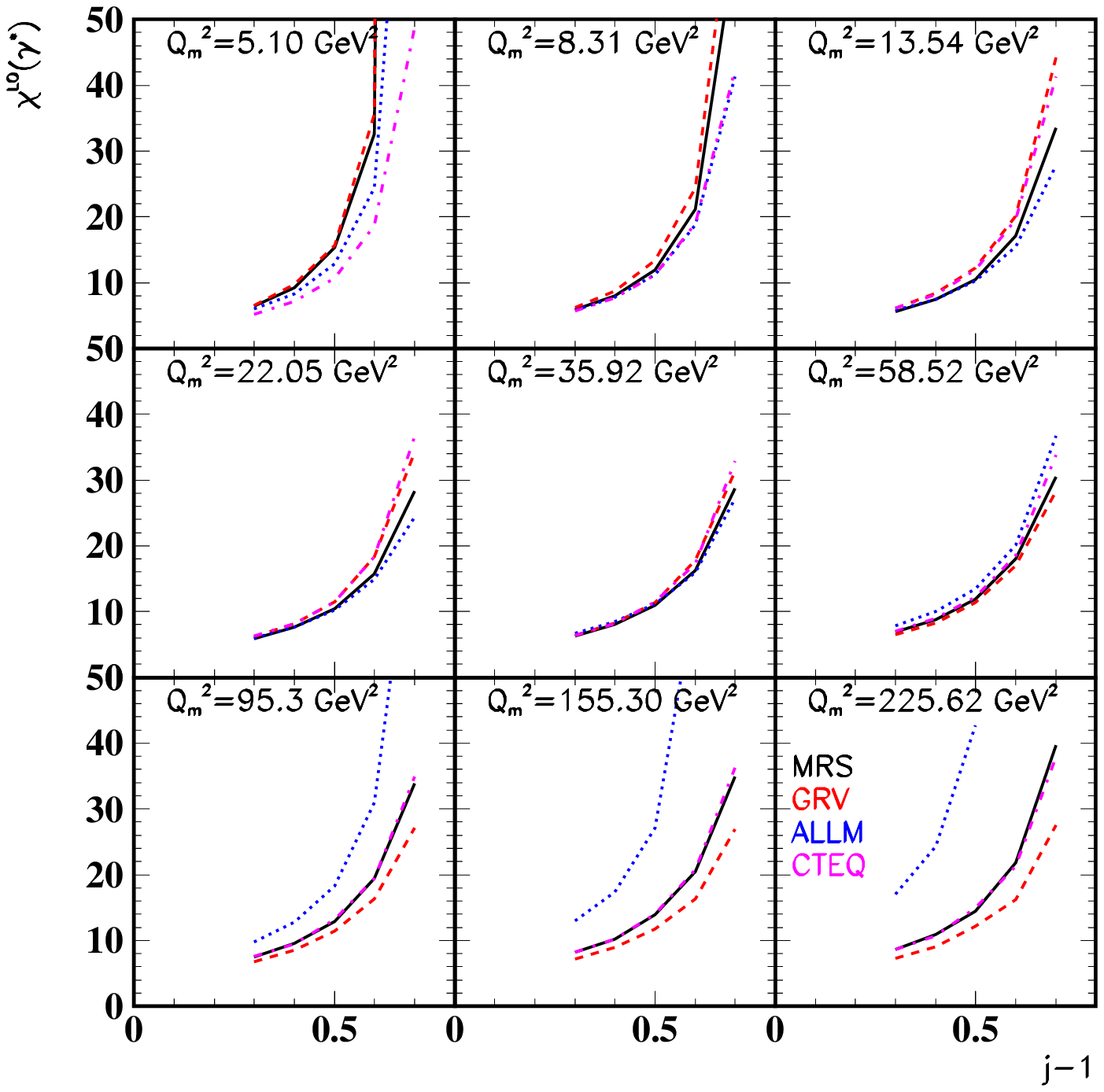} 
\vspace{-2cm} 
\caption{{\it Test of Relation {\bf ii)}: LO-BFKL.} }
\end{center}
\label{fig3}
\end{figure}
$\chi_{L0}$ is the  the LO-BFKL kernel,  
$\alpha_s$ the (fixed) coupling constant and 
$\omega( \gamma)$  an (unknown but factorizable\cite{impact}) non-perturbative 
coupling to the proton. 
\begin{figure}[ht]
\begin{center}
\epsfxsize=9cm   
\hspace{1.7cm}
\epsfbox{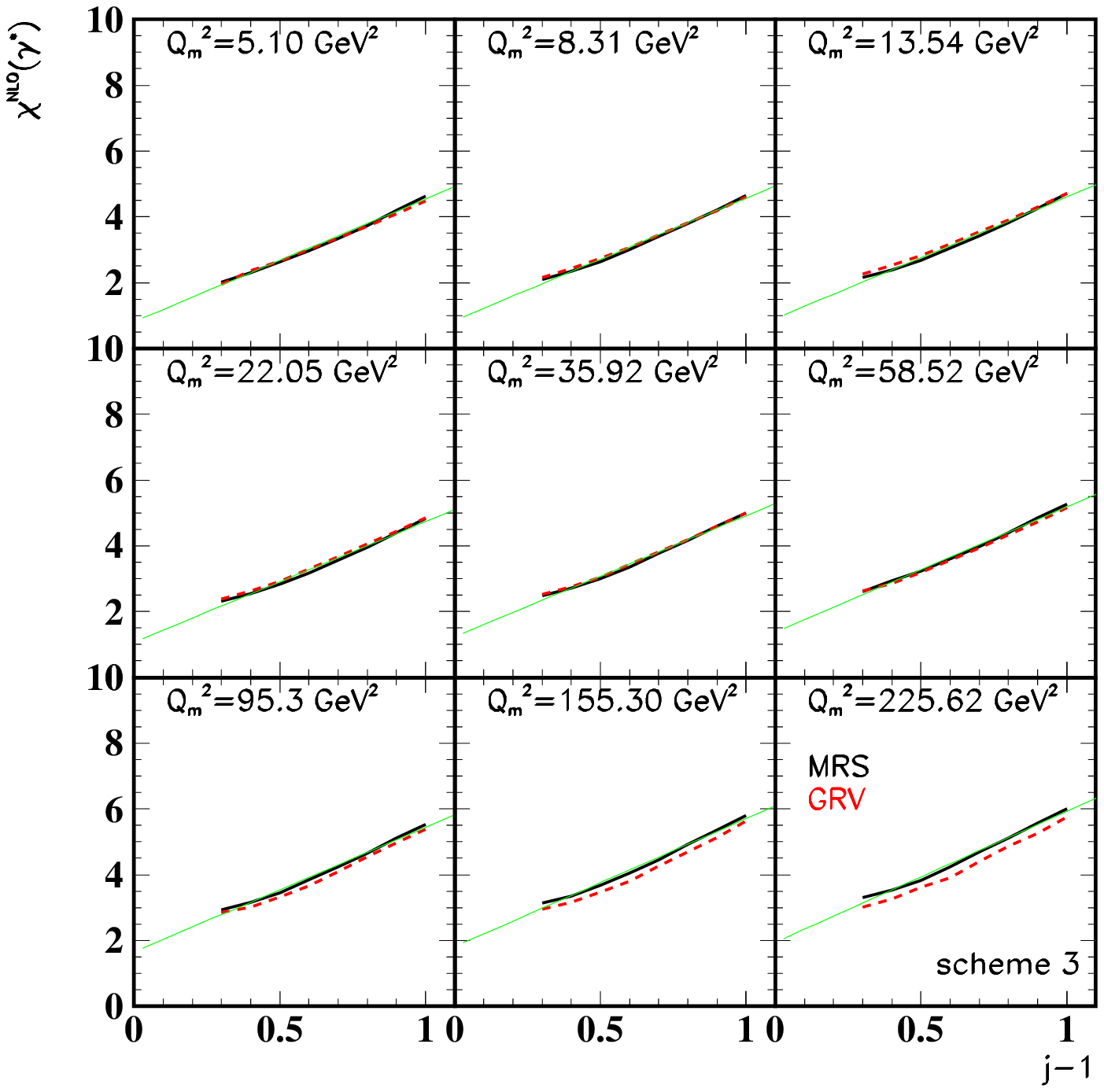} 
\vspace{-2cm} 
\caption{{\it Test of Relation {\bf ii)}: NLO-BFKL.} }
\end{center}
\label{fig4}
\end{figure}
Mellin-transforming (\ref{BFKL}) in $j\!-\!1\equiv \omega$ space, one easily 
finds
 \begin{equation}
\tilde F_i=\int 
\frac{d \gamma}{2i\pi}\left(\frac{Q^2}{Q_0^2}\right)^{ \gamma} 
\frac{1}{\omega-\frac{\alpha_s N_c}{\pi}\chi_{L0}( \gamma)}
\ h_i(\gamma)\ \eta( \gamma)\ ,
\label{mellin}
\end{equation}
and, taking the pole contribution, one has the  important relation 
 \begin{equation}
\omega=\frac{\alpha_s N_c}{\pi}\chi_{L0}( \gamma_k(\omega))\ .
\label{roots}
\end{equation}

Let us try and find the equivalent relation at NLO.  At (resummed) 
next-to-leading order, one can similarly 
write\footnote{Eq.(\ref{mellin-NLO}) is 
already an approximation of the (still) unknown complete (resummed) NLO-BFKL 
formula, since the photon and proton impact factors are not yet known at NLO. 
However, one  expects Eq.(\ref{mellin-NLO}) to be a phenomenologically valid 
approximation containing the information on the NLO kernel.}
 \begin{equation}
\tilde F_i=\int 
\frac{d \gamma}{2i\pi}\left(\frac{Q^2}{\Lambda_{QCD}^2}\right)^{ \gamma}
e^{-\frac {\ X(\gamma ,\omega)}{b \omega}}
\ h_i( \gamma,\omega)\ \eta( \gamma,\omega)\ ,
\label{mellin-NLO}
\end{equation}
where, by construction
 \begin {equation} 
{\partial \over \partial  \gamma } X( \gamma ,\omega)\equiv \chi_{NLO}( \gamma 
,\omega)\ .
\label{X}
 \end {equation} 
The function $X( \gamma ,\omega)$  appears in the solution
of  the Green function derived\footnote{The second variable of $X( \gamma 
,\omega)$ 
in 
(\ref{X}) corresponds to the choice of a reference scale $\mu\to \omega\equiv 
j\!-\!1$ dictated 
by the treatment of  the Green function fluctuations near the  
saddle-point\cite{salam}.} from the 
renormalization-group improved small-$x_{Bj}$ equation\cite{salam}, 
$\chi_{NLO}( \gamma ,\omega)$ is a resummed NLO-BFKL kernel and 
 \begin {equation} 
\frac {N_c}{\pi}\ 
\alpha_s(Q^2)=\left[{b\ln\left(Q^2/\Lambda_{QCD}^2\right)}\right]^{-1}\ ,
\label{al}
 \end {equation}
 with $b=11\!-\!2/3\ 
N_f/N_c.$

At large enough $Q^2/\Lambda_{QCD}^2,$ one can use the saddle-point 
appoximation 
in $\gamma$ to evaluate  (\ref{mellin-NLO}). Assuming that the perturbative 
impact factors 
 and the non-perturbative function $\eta$ do not vary much\footnote{We do not 
take into account modifications  {\it e.g.} coming from  powers of $\gamma$ in 
the 
prefactors which 
may shift the saddle point\cite{salam}. We thus assume a smoothness of the 
structure function integrand around the saddle-point in agreement with the 
phenomenology\cite{us}.},  the 
saddle-point condition reads
 \begin{equation}
\omega  \sim\frac {\chi_{NLO}(\bar\gamma,\omega)}{b\ 
\ln\left(Q^2/\Lambda_{QCD}^2\right)}\ 
 =
\frac{N_c\ \alpha_s(Q^2)}{\pi} \chi_{NLO}(\bar\gamma,\omega)\ ,
 \label{saddle}
 \end{equation}
where $\bar\gamma\equiv \bar\gamma(\omega,Q^2)$ is the saddle-point 
value. 

Inserting 
the saddle 
point defined by (\ref{saddle}) in formula (\ref{mellin-NLO}), one obtains a 
set of conditions to be fulfilled at (resummed) NLO level as follows:

{\bf i)}  The Mellin transform  $\tilde F_2 \equiv\tilde  F_T+\tilde F_L$ 
defines:
\begin{equation}
{\partial \over \partial \ln ({Q^2})} \ln \tilde F_2 (\omega,{Q^2}) \ \sim \ 
\bar 
\gamma(\omega,Q^2)\ . 
\label{i+)}
\end{equation}

{\bf ii)} $\bar \gamma$  verifies
\begin{equation}
\chi_{NLO}(\bar \gamma)\equiv \frac{\pi\ \omega}{\alpha_s (Q^2)
N_c} \ ,
\label{ii+)}
\end{equation}
where $\chi_{NL0}$ is a  resummed NLO-BFKL kernel.

{\bf iii)} The gluon structure function (one may also choose the obervable 
$F_L$)  verifies, {\it via} Mellin transform:
\begin{equation}
\ln(\tilde{G}(\omega,Q^2))=\ln\left(\tilde{F_2}(\omega,Q^2)\right)-\bar 
\gamma\ 
\ln\left[h_T(\bar \gamma)+
h_L(\bar \gamma)\right]\ .
\label{iii+)}
\end{equation}
We  test\cite{us} the  properties {\bf i)}-{\bf iii)} using NLO 
kernels 
proposed 
in\cite{salam}, and compared with  the LO kernel condition 
({\ref{roots}}).

As an example an extraction of an ``effective'' $F_2$ 
anomalous dimension {\bf i)}, see Fig. ({\ref{fig2}}), is 
performed\cite{us} using different parametrisations in a range of $\omega$   
verifying the stability 
with respect to  cuts on unknown  (smallest) or irrelevant (large) $x_{Bj}.$
The comparison of the property {\bf ii)} to the LO BFKL kernel is displayed in 
Fig.({\ref{fig3}}) and the one with a resummed NLO-BFKL kernels ({\it 
cf.} Scheme\cite{salam} 3) in Fig.({\ref{fig4}}). As is clearly seen 
on the 
figures the Mellin-transform analysis disfavors the BFKL-LO kernel, while it 
is 
qualitatively compatible with the resummed BFKL-NLO kernel. The remaining 
discrepancies at NLO could be attributed to finite NNLO 
corrections to the kernel or to still unknown NLO contributions to the impact 
factors\cite{bartels}. A systematic study of the proposed NLO kernels is thus 
made 
possible 
using the method\footnote{Similarly, relation  {\bf iii)} can be looked at 
using the 
gluon 
structure function parametrizations. However assumptions on the perturbative 
make the conclusions more 
qualitative or indicating some discrepancies  to be solved at NLO.}.

\section{ ``Saturation instability''}

As well-known, the BFKL evolution (even including next-leading contributions) 
leads 
to a multiplication of partons with non-vanishing size and thus inevitably 
leads 
to 
a dense medium . This may 
be  
called the ``Saturation Instability'' of the BFKL evolution.

The back-reaction of parton saturation on the BFKL equation has been 
originally\cite{GLR} described by adding a non-linear damping term. 
More 
recently, the evolution equation to saturation have been theoretically derived 
in 
the case of scattering on a ``large nucleus'', {\it e.g.} when the development 
of 
the parton cascade is tested by uncorrelated 
probes\cite{venugopalan,balitsky}. 

In the following we will focus on the solutions of the 
Balitsky-Kovchegov 
(BK) equations\cite{balitsky} where one consider the evolution within the QCD 
dipole Hilbert space\cite{mueller1}.
To be specific let us consider
$N(Y,{x}_{01}),$  the dipole forward 
scattering amplitude and define
\begin{equation}
{\mathbb N}(Y,k)=\int_0^{\infty} \frac{dx_{01}}{x_{01}}
J_0(kx_{01})\,N(Y,x_{01})\ .
\label{eq:fourier}
\end{equation}
\begin{figure}[ht]
\centerline{\epsfxsize=3.9in\epsfbox{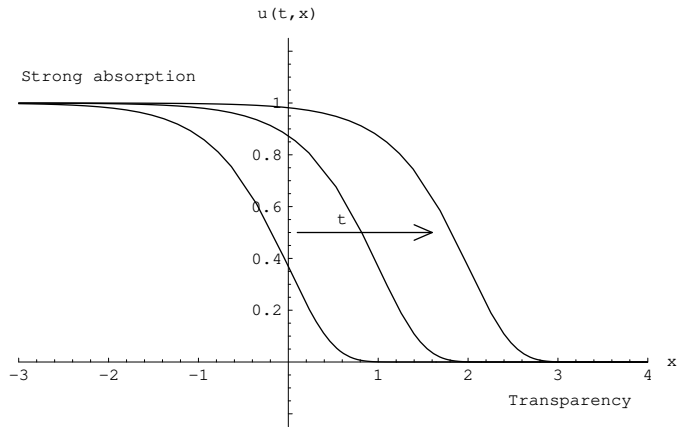}}   
\caption{{\it Typical traveling wave solution.} 
The function $u(t,x)$ is
represented for three different times. The wave front connecting the
regions
$u=1$ and $u=0$ travels
from the left
to the right as $t$ increases. That illustrates 
how the ``strong absorption'' or saturated phase
region invades the ``transparency'' region.
 \label{fig:fkpp}}
\end{figure}
Within suitable approximations
(large $N_c$, summation of fan diagrams, spatial homogeneity) 
and starting from the Balitsky-Kovchegov equation\cite{balitsky},
it has been shown  that this quantity obeys
the nonlinear evolution equation
\begin{equation}
{\partial_Y}{\mathbb N}=\bar\alpha
\chi\left(-\partial_L\right){\mathbb N}
-\bar\alpha\, {\mathbb N}^2\ ,
\label{eq:kov}
\end{equation}
where $\bar\alpha=\alpha_s N_c/\pi$,
$\chi(\gamma)=2\psi(1)-\psi(\gamma)-\psi(1\!-\!\gamma)$ is the
characteristic function of the BFKL kernel\cite{bfkl},
$L=\log (k^2/k_0^2)$ and $k_0$ is some fixed low momentum scale.
It is well-known that the BFKL kernel can be expanded 
to second order around $\gamma\!=\!{\scriptstyle \frac 12}$,
if one sticks to the kinematical regime $8\bar\alpha Y \gg L$. 
We expect this commonly used
approximation to remain valid 
for the full nonlinear equation.
The latter boils down to a parabolic nonlinear
partial derivative equation:
\begin{equation}
{\partial_Y}{\mathbb N}=\bar\alpha
\bar\chi\left(-\partial_L\right){\mathbb N}
-\bar\alpha\, {\mathbb N}^2\ ,
\label{eq:kovchegov}
\end{equation}
with
\begin{equation}
\bar\chi\left(-\partial_L\right)
=\chi\left({\scriptstyle \frac12}\right)+
\frac{\chi^{\prime\prime}\!\left({\scriptstyle\frac12}\right)}{2}
\left(\partial_L+{\scriptstyle \frac12}\right)^2\ .
\label{eq:expansion}
\end{equation}

The key point of our recent approach\cite{us1} is to remark that the structure 
of 
Eq.(\ref{eq:kovchegov}) is identical (for  fixed 
$\alpha$) to a mathematical  and physical archetype of non-linear evolution 
equation 
for 
which useful tools can be applied, namely the Fisher and 
Kolmogorov-Petrovsky-Piscounov (KPP) equation\cite{KPP} for a function $u$:
\begin{equation}
\partial_t u(t,x)=\partial_x^2 u(t,x)+u(t,x)(1-u(t,x))\ ,
\label{eq:KPP}
\end{equation} 
which is directly related\cite{us1} to ${\mathbb N}.$ The equation  can be 
generalized to many physical 
situations, including  running $\alpha.$

 \begin{figure}[ht]
\centerline{\epsfxsize=3.9in\epsfbox{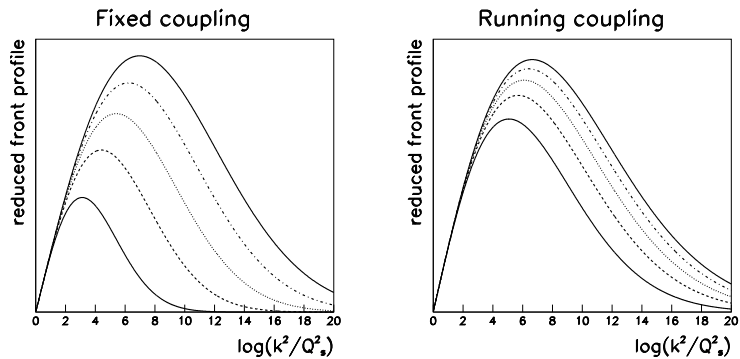}}   
\caption{{\it Evolution of the reduced front profile.} 
Fixed coupling: left; Running coupling: right.
The reduced front profile 
$(k^2/Q_s^2)^{\gamma_c}\,{\mathbb N}(k/Q_s(Y),Y)$ is plotted against
$\log(k^2/Q_s^2)$ for different rapidities.
 The various lines correspond to rapidities from 2 (lower
  curves, full line) up to 10 (upper curves).
Note the similarity of the wave fronts, but the quicker time evolution
(in $\sqrt{t}$) for fixed coupling, by contrast with the slow time
evolution (in $t^{1/3}$) for the running coupling case.
 \label{front}}
\end{figure}

Our main results are the following. The well-known geometric scaling 
property\cite{scaling} is obtained for the solution of the non-linear equation 
(\ref{eq:kovchegov}) for the gluon amplitude at large energy. 
In our notation, the geometric scaling property reads
\begin{equation}
{\mathbb N}(Y,k)={\mathbb N}\left(\frac{k^2}{Q_s^2(Y)}\right)\ ,
\label{eq:defscaling}
\end{equation}
where $Q_s^2(Y)$ is the saturation scale.
We prove that  geometric scaling is directly 
related to the existence of traveling wave solutions of the KPP 
equation\cite{KPP} at large times. This means that 
there exists a function 
of one variable $w$ such that 
\begin{equation}
u(t,x)\underset{t\rightarrow +\infty}{\sim}
w(x-m{(t)})
\end{equation}
uniformly in $x$.
Such a solution is depicted on Fig.(\ref{fig:fkpp}).
The function $m{(t)}$ depends on the initial condition. For the QCD 
case\cite{us1}, 
one has to consider
\begin{equation}
m{(t)}=2 t-{\scriptstyle \frac32}\log t+{\mathbb O}(1)\ .
\label{eq:velocity}
\end{equation}

When transcribed in the appropriate physical variables, this mathematical 
result, implies directly the known geometric scaling properties\cite{us1}. It 
is interesting to note how the traveling wave solutions provide a particularly 
striking mathematical realization of an ``instability'' as depicted in 
Fig.(\ref{fig:fkpp}), when a stable fixed point (strong absorption) 
``invades'' 
an unstable one (transparency).

This mathematical analysis can be extended\cite{us1} to the study of the 
transition towards geometrical scaling , {\it i.e.} the formation of the front 
wave as a function of time,  both for fixed and running $\alpha,$ see 
Fig.(\ref{front}).

\section{Conclusion}

In the present contribution, we have discussed some aspects of the 
``instabilities'' of the BFKL equations, {\it i.e.}:

 \begin{itemize}
\item 
{\bf i)} {\it Instability towards the non-perturbative regime}

\item 
{\bf ii)} {\it Instability towards the renormalization group  regime}

\item 
{\bf iii)} {\it Instability towards the high density (saturation) regime}

\end{itemize}

At first sight, these  instabilities could have appeared   as  drawbacks of 
the whole 
approach. 
On the very contrary, we have seen that the extensions of 
BFKL equation raised up by the treatment of  ``instabilities'' appear to be  
the building blocks of  most interesting recent 
developments 
towards 
a better understanding of  QCD dynamics. As an example, I chose to present 
some personal recent contributions to this discussion, which are far from 
giving an idea of the whole extent of the works\footnote{I  mentioned quite a 
few of  them in the reference list but I want to apologize for the authors and 
studies which I may have forgotten in this necessarily shortened review.} 
which attack the problem nowadays.

As a brief outlook, let us mention:

About Point
({\bf i)}, not discussed here, let us mention  the formal but informative 
discussion on the $N=4$ supersymmetric QCD field theory and the AdS/CFT 
correspondence\cite{strings}. 

Point ({\bf ii)}: It is the subject of a developing activity which will allow 
to master the rather high technicality of the BFKL-NLO calculations and thus 
to penetrate the subtle aspects of the compatibility between BFKL and DGLAP 
evolution equations.

Point ({\bf iii)}: Saturation with both its phenomenological and theoretical  
aspects will certainly retain the attention of the Particle Physics community. 
The challenge here is the quest for a new phase of intense QCD fields and the 
undersatanding of its dynamical properties.

\section*{Acknowledgments}
I want to warmly thank my collaborators in the work which has been discussed 
here: 
St\'ephane Munier, Christophe Royon, Laurent Schoeffel and many colleagues 
with 
whom I had fructuous discussions, including those taking place in the charming 
decor of 
Ringberg Castle, in front of the Bavarian Alps.

\end{document}